\documentclass[twoside,fleqn]{article}
\usepackage{espcrc2,latexsym}
\usepackage{amsmath}
\usepackage{graphicx}
\usepackage{amsfonts}
\usepackage{amssymb}
\usepackage{epsfig}% one of graphics,graphicx,epsfig

\title{Dynamical Fermions in Hamiltonian Lattice Gauge Theory}
\author{Dean Lee
\address{Dept. of Physics, North Carolina State University, Raleigh, NC 27695}}

\begin{document}

\begin{abstract}
We describe a first attempt to understand dynamical fermions within a
Hamiltonian framework. \ As a testing ground we study compact QED$_{3}$, which
shares some important features of QCD$_{4}$ such as confinement, glueballs,
mesons, and chiral symmetry breaking. \ We discuss the methods used and show
data for the chiral condensate.
\vspace*{-1mm}
\end{abstract}
\maketitle

\section{Introduction}

Nearly all lattice studies with dynamical fermions rely on pseudofermion
techniques. \ This includes the standard Hybrid Monte Carlo method for
non-local boson actions \cite{hmc} and more recently multiboson methods
proposed by L\"{u}scher \cite{luscher} and extended to odd flavors by
Bori\c{c}i and de Forcrand \cite{borici}. While these methods are presently
the most efficient available for dynamical fermion calculations, much
improvement is still needed to combat algorithmic slowdown at light quark
masses and large volumes. \ On the more theoretical side, one would also like
to understand the microscopic dynamics of quarks in lattice simulations.
\ Ideally one hopes to see the effects of confinement on quark-antiquark pair
separation and the effects of chiral symmetry breaking on light quark
mobility. \ These questions are most easily answered in a framework which does
not integrate out the fermionic degrees of freedom. \ With these goals in mind
we investigate the possibility of simulating dynamical fermions in a confining
gauge theory with explicit fermion dynamics.

In this proceedings article we describe the beginning of a research program at
NC State to understand dynamical fermions within a Hamiltonian framework. \ As
a testing ground we study compact QED$_{3}$, which 
exhibits confinement, glueballs, mesons, and chiral
symmetry breaking. \ We discuss the methods that are being used and show
preliminary data for the chiral condensate\ and compare with strong and weak
coupling results.

\section{Compact QED$_{3}$}

We follow the notation used in \cite{burden} and \cite{hamer}. \ The
Hamiltonian is
\begin{equation}
H=\tfrac{1}{2a\sqrt{x}}W\text{,}%
\end{equation}
where $a$ is the lattice spacing and $x$ the dimensionless strong coupling
constant,
\begin{equation}
x=e^{-4}a^{-2}\text{.}%
\end{equation}
$W$ has three pieces,
\begin{equation}
W=W_{0}+\sqrt{x}W_{1}+xW_{2}\text{,}%
\end{equation}
where
\begin{align}
W_{0}  & =\sum_{l}E_{l}^{2}-\mu\sum_{\vec{r}}(-1)^{r_{1}+r_{2}}\chi^{\dagger}(\vec{r})\chi(\vec{r})\\
W_{1}  & =\sum_{\vec{r},j}\eta_{j}(\vec{r})\left[  \chi^{\dagger}(\vec{r})U_{j}%
(\vec{r})\chi(\vec{r}+\hat{\jmath})+\text{h.c.}\right] \nonumber\\
W_{2}  & =-\sum_{p}\left(  U_{p}+U_{p}^{\dagger}\right)  \nonumber
\end{align}
and $\eta_{1}(\vec{r})=(-1)^{r_{2}+1}$, $\eta_{2}(\vec{r})=1$.  We use four component staggered fermions, which in the three-dimensional
Hamiltonian formalism gives one flavor of fermion. \ The dimensionless mass
parameter $\mu$ is given by
\begin{equation}
\mu=2me^{-2}\text{.}%
\end{equation}

\section{Methods}

We start with a set of trial states $\left|  \Psi_{j}\right\rangle$
and evaulate the quantities
\begin{align}
M_{ij}  & =\left\langle \Psi_{i}\right|  \exp\left[  -HT\right]  \left|
\Psi_{j}\right\rangle ,\\
M_{ij}^{\Delta}  & =\left\langle \Psi_{i}\right|  \exp\left[  -H(T+\Delta
T)\right]  \left|  \Psi_{j}\right\rangle .\nonumber
\end{align}
Improved trial states are given by
\begin{equation}
\left|  \Psi_{j}^{\prime}\right\rangle =\exp[-\tfrac{1}{2}HT]\left|  \Psi
_{j}\right\rangle .
\end{equation}
It is easy to see that
\begin{align}
M_{ij}  & =\left\langle \Psi_{j}^{\prime}\right.  \left|  \Psi_{j}^{\prime
}\right\rangle ,\\
M_{ij}^{\Delta}  & =\left\langle \Psi_{j}^{\prime}\right|  \exp\left[  -H\Delta
T\right]  \left|  \Psi_{j}^{\prime}\right\rangle .
\end{align}
We now find a matrix $T$ and unitary matrix $U$ such that
\begin{align}
T_{ij}^{\dagger}M_{jk}T_{kl}  & =\delta_{il},\\
U_{ij}^{\dagger}T_{jk}^{\dagger}M_{kl}^{\Delta}T_{lm}U_{mn}  & =\lambda
_{i}\delta_{in}.\nonumber
\end{align}
From these we can read off approximate eigenvectors
\begin{equation}
\left|  E_{l}\right\rangle =T_{lm}U_{mj}\left|  \Psi_{j}^{\prime
}\right\rangle
\end{equation}
and eigenvalues
\begin{equation}
\left|  E_{l}\right\rangle =-\tfrac{1}{\Delta T}\log\lambda_{l}\text{.}%
\end{equation}

The matrices $M$ and $M^{\Delta}$ are evaluated by dividing into time slices,
\begin{equation}
M_{ij}=\left\langle \Psi_{j}^{\prime}\right|  \exp\left[  -\tfrac{HT}%
{N}\right]  \cdots\exp\left[  -\tfrac
{HT}{N}\right]  \left|  \Psi_{j}^{\prime}\right\rangle \text{.}%
\end{equation}
Between the time slices we insert a complete set of states, and the basis we
choose is the tensor product basis of gauge and fermion states. \ For the
gauge states we use coherent states of the link fields,%

\begin{align}
U_{j}(\vec{x})
%TCIMACRO{\tbigotimes_{l,\vec{r}}}%
%BeginExpansion
{\textstyle\bigotimes_{l,\vec{r}}}
%EndExpansion
\left|  A_{l}(\vec{r})\right\rangle   & =e^{iA_{j}(\vec{x})}
%TCIMACRO{\tbigotimes _{l,\vec{r}}}%
%BeginExpansion
{\textstyle\bigotimes_{l,\vec{r}}}
%EndExpansion
\left|  A_{l}(\vec{r})\right\rangle \\
U_{j}^{\dagger}(\vec{x})
%TCIMACRO{\tbigotimes _{l,\vec{r}}}%
%BeginExpansion
{\textstyle\bigotimes_{l,\vec{r}}}
%EndExpansion
\left|  A_{l}(\vec{r})\right\rangle   & =e^{-iA_{j}(\vec{x}%
)}
%TCIMACRO{\tbigotimes _{l,\vec{r}}}%
%BeginExpansion
{\textstyle\bigotimes_{l,\vec{r}}}
%EndExpansion
\left|  A_{l}(\vec{r})\right\rangle  .\nonumber
\end{align}
For the fermion states we use the usual position-space basis. \ For fixed
coherent states of the gauge field between time slices, the fermion
dynamics are generated a classical background field,
\begin{equation}
\chi^{\dagger}(\vec{r})U_{j}(\vec{r})\chi(\vec{r}+\hat{\jmath})\rightarrow
e^{iA_{j}(\vec{r})}\chi^{\dagger}(\vec{r})\chi(\vec{r}+\hat{\jmath}).
\end{equation}

\section{Simulations}

In our calculations we keep the lowest few energy fermionic states in computer
memory and calculate their dynamics exactly by matrix multiplication. \ We use
a simple heat bath algorithm to propose new gauge updates. \ These updates are
performed by sweeps through the spatial lattice, taking each time slice in
succession. \ We use a Metropolis accept/reject decision based upon an
estimator that includes the contribution of the low energy fermion states.
\ In comparison with quenched simulations, the use of this estimator is slower
by a constant factor proportional to the number of low energy fermion states. \ After each sweep through all time slices, the contribution
of other fermionic states are calculated using diffusion Monte Carlo sampling.

In our initial studies of small to mid-sized lattices we store in memory the nine
lowest energy fermion states in memory, the strong coupling ground state and
the eight translationally invariant states connected to the ground state by a
single hop. \ In Figure 1 we show results for $\mu=2$. \ The data is
extrapolated from spatial lattice sizes $4\times4$, $8\times8$, and
$10\times10$, and number of time slices $11$, $15$, and $19$. \ The curves
$s_{0},s_{1},s_{2},s_{3}$ give the first few orders in the strong coupling
expansion, and $w$ is the leading behavior in the weak coupling expansion.
\ We find behavior consistent with strong coupling estimates at small $x$ and
weak coupling estimate at large $x$.

\section{Conclusions}

We have described a first attempt to treat dynamical fermions within a
Hamiltonian framework. \ We considered compact QED$_{3}$, which shares some
important features of QCD$_{4}$. \ At this time it is too early to say whether
this approach would be viable for QCD$_{4}$ with relatively light quarks. \ We
are currently making accuracy and efficiency comparisons with pseudofermion
methods in QED$_{3}$ \cite{burkitt} for exactly massless fermions.

On the theoretical side there is clearly much potential in this approach to
better understand the microscopic dynamics of quarks in lattice simulations. \ In
future work we plan to study the effects of confinement and chiral symmetry
breaking on quark dynamics.\medskip

\it
The author thanks Pieter Maris and Nathan Salwen for discussions.
\rm

\begin{figure}[t]
\epsfig{file=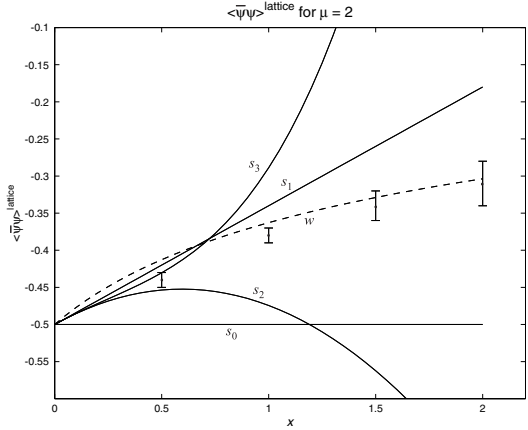,width=7cm}
\caption{Chiral condensate for $\mu=2$. $s_{0,}s_{1},s_{2},s_{3}$ represent
the lowest order estimates in the strong coupling expansion, and $w $ is the
leading result in the weak coupling expansion.}%
\end{figure}

\end{document}